\documentclass[12pt]{article}
\usepackage{graphicx}
\usepackage[margin=1in]{geometry}
\usepackage[final,nopatch=footnote]{microtype}

\usepackage{xr-hyper}
\usepackage[breaklinks, colorlinks=true, urlcolor=blue, citecolor=blue, linkcolor=blue]{hyperref}
\usepackage{natbib}
\usepackage{setspace}
\usepackage{mathptmx}
\usepackage[flushleft]{threeparttablex}
\usepackage{booktabs}
\usepackage{wrapfig}
\usepackage{listings}
\usepackage{amsmath}
\usepackage[hang,flushmargin]{footmisc}
\usepackage[T1]{fontenc}
\usepackage{longtable}
\usepackage{caption}

\AtBeginEnvironment{thebibliography}{\singlespacing\small}
\setcitestyle{authoryear,open={(},close={)}}

\newcommand{\numberthis}{\addtocounter{equation}{1}\tag{\theequation}}
\newcommand{\mytilde}{\raise.17ex\hbox{$\scriptstyle\sim$}}

\title{
    \Large Personality pairing improves human--AI collaboration
}
\author{
    \vspace{-1em} \small Harang Ju\\
    \vspace{-1em} \small Carey Business School, Johns Hopkins University \\
    \vspace{-1em} \small harang@jhu.edu
    \and
    \vspace{-1em} \small Sinan Aral\\
    \vspace{-1em} \small Sloan School of Management, Massachusetts Institute of Technology \\
    \small sinan@mit.edu
}
\date{}

\begin{document}

\maketitle

\begin{abstract}
    \onehalfspacing
     Here we examine how AI agent ``personalities'' interact with human personalities to shape human--AI collaboration and performance. In a large-scale, preregistered randomized experiment, we paired 1,258 participants with AI agents prompted to exhibit varying levels of the Big Five personality traits. These human--AI teams produced 7,266 display ads for a real think tank, which we evaluated using 1,168 independent human raters, and a field experiment on X that generated nearly 5 million impressions. We found that human and AI personalities individually shaped ad quality and teamwork and that human--AI personality pairings directly influenced ad quality. For example, extraverted humans paired with conscientious AI produced the lowest quality ads, followed by conscientious humans paired with agreeable AI and neurotic humans paired with conscientious AI. In the field experiment, ad quality significantly influenced ad performance, measured by click-through rates and cost-per-click. Together, these results demonstrate that personality pairing can improve human--AI collaboration and performance. They also motivate future research on the complex implications of AI personalization for human--AI collaboration, teamwork, and performance.
\end{abstract}

\clearpage

AI tools have gained significant attention for their ability to increase productivity and performance across a wide range of tasks. For example, chatbots powered by large language models (LLMs) reduced the time required for mid-level writing tasks by 40\% while improving quality by 18\% \citep{noy2023experimental}. AI-enhanced resumes increased job hiring rates by 8\% \citep{wiles2023writing}. In customer support settings, AI assistance increased productivity by an average of 14\% \citep{brynjolfsson2023genai}. But these benefits also exhibited considerable heterogeneity, varying across tasks \citep{dellacqua2023navigating, ju2025collaborating}, worker seniority \citep{wang2023friend}, gender \citep{dellacqua2023navigating, humlum2024adoption}, and skill levels \citep{noy2023experimental, brynjolfsson2023genai, otis2024uneven}. In clinical settings, AI assistance reduced reading time while maintaining diagnostic accuracy, but effects varied significantly across physicians and clinical contexts \citep{chen2024impact, yu2024heterogeneity, agarwal2023combining}. In financial services and customer support, human involvement in AI-based services boosted consumer confidence and material outcomes, though in other contexts AI reduced quality due to sycophancy and overreliance \citep{yang2025my, vaccaro2024combinations, sergeyuk2025human}. Unfortunately, experiments documenting these productivity and performance gains have generally provided participants with identical AI tools, leaving open the possibility that at least some of the observed variation in performance stems from a mismatch between AI and individual human characteristics. Performance gains may therefore increase and simultaneously vary less by age, gender, or skill level, if AI is personalized for each human collaborator.

Such personalization may now be possible as foundation model companies have recently begun developing, refining, and deploying AI personalities to improve human--AI interaction. For example, OpenAI introduced personality customization in models like GPT-5, offering Default, Cynic, Robot, Listener, and Nerd personalities \cite{openai2025sycophancy, openai2025customizing}. Similarly, Anthropic has begun to measure undesirable personality shifts during training, such as sycophancy, a tendency for AI to overly align with user preferences \cite{anthropic2024character, sharma2025sycophancy}, and introduced novel training methods to mitigate such personality shifts in their models. These advances enable AI agents to be tailored to users, potentially improving human--AI collaboration and ushering in a new era of AI personalization.

Despite the development of AI personalities, no prior research has rigorously and experimentally examined their causal effects on human--AI collaboration. Such experiments are now critical to understanding whether and how tailored AI personalities may impact teamwork and performance as the lack of experimental evidence on AI personalization hinders the optimization of AI systems for collaborative tasks and undermines our understanding of the detrimental effects of mismatched AI personalities on, for example, adverse psychosocial outcomes \citep{fang2025psychosocial}.

In human teams, team composition is a critical factor that promotes trust, coordination, and teamwork \citep{zhang2023teammate, hemmer2023, woolley2023collective}. People exhibit variation in demographic, psychological, and behavioral traits known to impact team performance \citep{kichuk1997, peeters2006, woolley2010ci}. Theories of person-team fit suggest that when team members' personalities are compatible, collaboration becomes more effective and leads to better results \citep{barrick1998gma, bell2007meta}. As AI has evolved from a passive tool to an active collaborator \citep{makarius2020rising, fugener2022cognitive, anthony2023, collins2024building, ju2025collaborating}, tools like Cursor or Claude Code now act on behalf of and co-create alongside human users, as team members, in real time. Like humans, these AI agents also possess characteristics that influence teamwork, technology acceptance, and even psychosocial outcomes \citep{li2025acceptance, tully2025lower, fang2025psychosocial}.

Early work on computers as social actors showed that people prefer personality-matched computers \citep{nass2000machines}, and personality-targeted advertising can alter behavior at scale \citep{matz2017psychological, matz2024persuasion}. However, these studies examined either preference outcomes with pre-LLM systems or one-shot content personalization, not interactive collaboration with AI agents. The interaction between human and AI traits in collaborative production remains understudied \cite{krakowski2025human}. With AI agents increasingly operating as social actors in collaborative workflows \citep{anthony2023, gelbrich2025}, it is crucial to understand whether the principles of personality pairing apply to human--AI teams and whether AI agents can be designed to exhibit personalities that complement those of their human counterparts.

To address these gaps in our understanding of human--AI collaboration, we examined how AI personalities interact with human characteristics to shape teamwork and performance. Our study leverages the unique capabilities of AI agents to vary their individual behavioral styles \citep{jiang2023evaluating, chen2025persona}, allowing us to investigate how human--AI personality pairing influences the performance of human--AI teams. Systematic prompt variation can effectively shape AI agent behavior \citep{schulhoff2024prompt}, and seminal work on OpenAI's ChatGPT architecture demonstrates that prompting enables control of agent behavior without task-specific fine-tuning \citep{brown2020language}. Advanced prompting methods, such as chain-of-thought prompting, have further demonstrated significant improvements in task performance by guiding a model's reasoning processes \citep{wei2023chainofthought} and ``persona vectors'' have been shown to successfully monitor and control personality shifts in AI agents \citep{chen2025persona}. These methods enable us to intentionally design AI agents with particular social and behavioral characteristics, which can be tailored to complement human collaborators. Randomized controlled trials (RCTs) are now needed to evaluate the effects of such AI personalities on human--AI collaboration and task outcomes to understand how AI agents can be personalized for effective teamwork \citep{jakesh2023opinionated}.

Investigating human--AI collaboration with agentic tools is challenging in practice. These tools must be seamlessly embedded in workspaces, have full contextual awareness, and be capable of taking actions on behalf of users. Developing such sophisticated tools is a complex and costly endeavor, and rigorous research in this area requires multiple treatment arms to test interactions with individual covariates, necessitating large sample sizes to ensure statistical power. Downstream performance measures are also crucial to evaluating the effectiveness of real collaborations in the field, not just in the lab. Our study addresses all of these requirements, providing a comprehensive framework for investigating the real-world economic impact of AI personalization.

Here, we found that personality pairing shaped outcomes from the lab to the field. In the lab, we measured pairing effects of up to two-fifths of a SD on human-rated ad quality. We then traced that quality into the field, where a one-standard-deviation increase in human-rated text quality predicted a 7.0\% higher click-through rate and the same increase in image quality predicted a 3.9\% lower cost-per-click, or \$0.30. Taken together, our results show that even a single line of personality prompting, applied upstream of the collaboration, can shift both the quality of the work and its market performance.

\section*{Experiment Design}

To investigate how personality pairing between humans and AI agents influences collaboration, we conducted a large-scale experiment involving 1,258 participants representative of the US population using Pairit, an experimental platform we developed for extensible human--AI collaboration \citep{ju2025collaborating}. After taking a 10-item pretask survey to measure their Big Five personality traits, participants, who were recruited via Prolific from October 15 to 18, 2024, were randomly paired with multimodal AI agents, which had full context of the workspace and were induced to independently exhibit high or low levels of the Big Five personality traits using P\textsuperscript{2} prompting \citep{jiang2023evaluating}. These human--AI teams then worked together to create display ads marketing the year-end report of a real think tank during a 40-min session, using real-time chat and synchronized text and image editing tools. Teams were provided with image assets by the think tank and could separately call Dall-E 3 to generate new images for the campaigns.

Throughout the collaboration, the AI agent dynamically received screenshots of the current ad image alongside the task brief, the current ad copy, the chat history with the human, and its own action history. As it was provided this context, it could choose from five action types: Wait, Chat, EditText (revise the headline, primary text, description, or image prompt), SelectImage (choose an image from the carousel), or GenerateImage (call Dall-E~3 to produce a new image). The agent and the human shared a single live workspace, so the human could see and edit the same ad in real time as the agent acted. Only the human partner could submit ads, and teams iterated to submit as many ads as they could within the 40-min window. This setup allowed us to assess the causal effects of personality pairing on collaboration outcomes. The randomization of AI agent personalities, such as high or low conscientiousness, openness, or extraversion, enabled us to isolate the causal effects of human--AI personality pairing on the collaboration and performance of human--AI teams.

Importantly, we approached personality prompts as neither about memory, such as recalling past chat interactions, nor about aligning LLM responses to human preferences which often leads to AI sycophancy \citep{chhikara2025mem0, sharma2025sycophancy}. Instead, we treated AI personality as an experimental treatment: Each AI agent was independently and randomly assigned a personality profile (high or low on each Big Five personality trait), and we estimated how this assignment interacted with the human partner's measured traits to shape collaboration outcomes. Random assignment rules out selection effects, such as users gravitating toward AI styles they prefer, so observed differences across pairings reflect causal effects of the pairing itself. This approach parallels findings in human--robot interaction, where robots' social communication and appearance have been shown to influence team performance \citep{jung2013backchanneling, jung2018theories}.

To assess the quality of the display ads produced by the human--AI teams, we conducted a separate evaluation study using a second independent sample of 1,168 new participants recruited from Prolific. These raters were based in the United States and selected to ensure representation across gender, age, and ethnicity. Each rater evaluated a random sample of 40 ads, with the sampling design guaranteeing that every ad received at least three independent ratings. Throughout the paper, ad quality refers to these human-rated evaluations of text quality, image quality, and click likelihood. Raters assessed each ad on three dimensions using 7-point Likert scales: text quality (``The text is present, clear, relevant, and engaging''), image quality (``The image is visually appealing''), and likelihood of clicking the ad (``I am likely to click on this ad''). The ads were presented as realistic mockups, incorporating standard display ad elements such as images, copy, call-to-action buttons, and social media--style interaction options (e.g., Like, Comment, Share; see Materials and Methods).

To evaluate the real-world performance of the display ads, we then conducted a preregistered field experiment on the X platform. We sampled 1,152 ads from the 7,266 human--AI ads produced during the experiment, using stratified random sampling based on the predicted click likelihoods implied by the human ratings to ensure sampling across the ad quality distribution. These ads were then run in 400 geographically nonoverlapping campaigns, each with a five-ad split test \citep{braun2024abtesting, braun2024adplatforms}. Key performance metrics were the click-through rate (CTR) and the cost-per-click (CPC) measured on X, along with the view-through rate (VTR) and view-through duration (VTD) tracked via unique DocSend links for each ad (details of the field experiment are provided in Materials and Methods). We employed ZIP-code-level randomization across campaigns to prevent audience spillovers, and campaign random effects absorbed campaign-level variation in audience composition and temporal confounders. Because advertising spend is partly determined by autobidding after deployment, the quality--performance regressions adjust for realized advertising spend. We applied Benjamini--Hochberg (BH) multiple hypothesis testing corrections to all analyses in the paper and report results based on these corrections throughout.

\section*{Results}

Our experimental results demonstrate that personality pairing affected human-rated ad quality, teamwork, and real-world ad performance. We first report the individual effects of human and AI personality traits on teamwork and ad quality, then turn to the personality pairing results and conclude by reporting the results of the field experiment.

\subsection*{Individual Trait Effects on Teamwork and Ad Quality}

\begin{figure}[ht]
\centering
\includegraphics[width=0.7\linewidth]{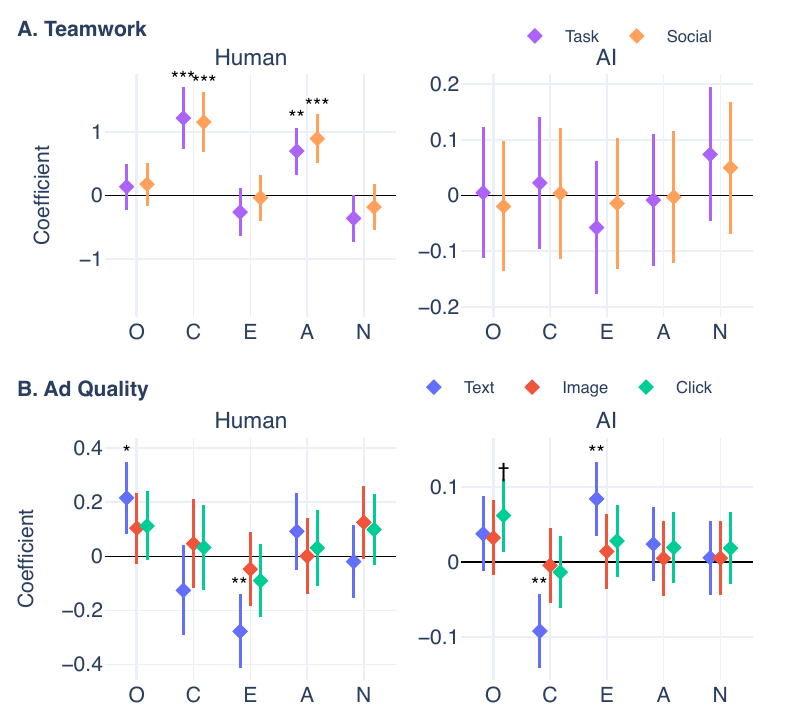}
\caption{Effect of personality on ad quality and teamwork. Coefficients from regressions of each outcome on the five human trait scores and the five AI trait scores, with no interaction terms, BH-corrected within each outcome family. Panel (A) shows teamwork indices (task, social); panel (B) shows ad quality dimensions (text, image, click likelihood). Within each panel, human trait effects appear on the Left and AI trait effects on the Right. X-axis labels use O = Openness, C = Conscientiousness, E = Extraversion, A = Agreeableness, and N = Neuroticism. Symbols indicate significance after BH correction: $^{\dagger}P < 0.10$, $^{*}P < 0.05$, $^{**}P < 0.01$, and $^{***}P < 0.001$. Error bars show 95\% CIs. See SI Appendix, section S7, for the full regression tables.}
\label{fig:main_effects}
\end{figure}

We first tested whether measured human traits and randomized AI traits independently predicted teamwork and ad quality in human--AI teams (Fig.~\ref{fig:main_effects}). Each model regressed an outcome on the five human trait scores and the five AI trait scores with no interaction terms, and BH corrections were applied within each outcome family. Since AI traits were experimentally randomized, while human traits were measured but not manipulated, the AI coefficients can be interpreted causally, whereas the human trait coefficients represent correlations not causal relationships.

Human traits predicted both teamwork and ad quality. Conscientious and agreeable humans experienced stronger task ($P < 0.001$, $P < 0.01$) and social teamwork ($P < 0.001$, $P < 0.001$). Conscientiousness and agreeableness in humans also predicted higher ratings on all six underlying teamwork dimensions (communication, coordination, balance of contributions, mutual support, effort, and cohesion) (SI Appendix, section S7D). Human openness predicted higher text quality ($P < 0.05$) and human extraversion predicted lower text quality ($P < 0.05$). AI traits also affected ad quality: AI extraversion increased text quality ($P < 0.05$), whereas AI conscientiousness reduced text quality ($P < 0.01$) (see SI Appendix, section S7D, for the full regression tables). It is perhaps not surprising that human and AI personalities affect teamwork and ad quality. But evaluating the potential benefits of AI personalization requires a more comprehensive analysis of human--AI personality pairings.

\begin{figure}[h]
\centering
\includegraphics[width=\textwidth]{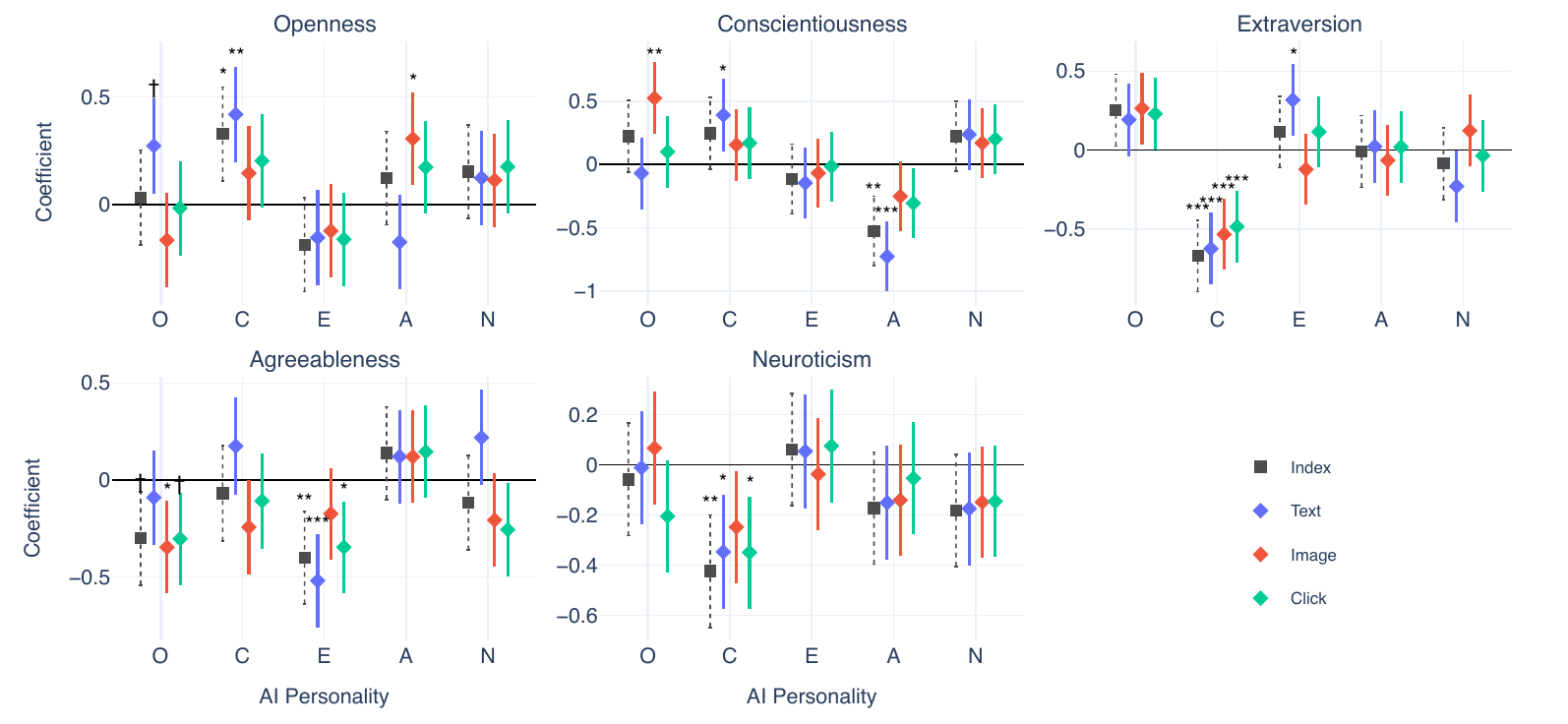}
\caption{Effect of personality pairing on ad quality. Coefficients for human--AI personality interaction terms on the composite ad quality index (gray) and the three ad quality dimensions: text quality (blue), image quality (red), and click likelihood (green), BH-corrected within each outcome family. Panel titles show the full human trait names; x-axis labels use O = Openness, C = Conscientiousness, E = Extraversion, A = Agreeableness, and N = Neuroticism. Symbols indicate significance after BH correction: $^{\dagger}P < 0.10$, $^{*}P < 0.05$, $^{**}P < 0.01$, and $^{***}P < 0.001$. Error bars show 95\% CIs. See SI Appendix, section S7, for the full regression table.}
\label{fig:personality}
\end{figure}

\subsection*{Personality Pairing and Ad Quality}

Beyond individual traits, personality pairings between humans and AI agents significantly influenced ad quality (Fig.~\ref{fig:personality} and SI Appendix, section S17). Fig.~\ref{fig:personality} reports the full set of pairing estimates. The largest three effects on the composite human-rated ad-quality score were for extraverted humans paired with conscientious AI ($\beta = -0.670$, $p_{\mathrm{BH}} < 0.001$), conscientious humans paired with agreeable AI ($\beta = -0.525$, $p_{\mathrm{BH}} = 0.002$), and neurotic humans paired with conscientious AI ($\beta = -0.424$, $p_{\mathrm{BH}} = 0.002$). These pairings reduced ad quality by 0.39, 0.29, and 0.26 SD, respectively, on the 1 to 7 composite score. For reference, the composite score has a mean of 4.35 and a SD of 1.47 in our sample. Notably, a single line of personality prompting produced each of these changes in ad quality, well upstream of the full collaborative production process.

At the level of individual performance indicators, extraverted humans paired with conscientious AI lowered quality across all three measured dimensions including text ($P < 0.001$), image ($P < 0.001$), and click likelihood ($P < 0.05$), while pairing with extraverted AI increased text quality ($P < 0.01$). Conscientious humans produced lower quality text when paired with agreeable AI ($P < 0.001$) and higher quality text and images when paired with conscientious ($P < 0.01$) and open AI ($P < 0.05$). Open humans produced higher quality text when paired with conscientious AI ($P < 0.05$) and lower quality text when paired with agreeable AI ($P < 0.05$). Agreeable humans produced lower quality text when paired with extraverted AI ($P < 0.05$) and lower quality images when paired with open AI ($P < 0.05$). Finally, neurotic humans produced ads with lower click likelihood when paired with conscientious AI ($P < 0.05$).

To understand how these effects emerged in the collaboration, we labeled the human--AI chat logs with an LLM, focusing on the three largest pairing effects. In each, the AI questioned or pushed back slightly less and took direction slightly more than the same AI prompted low on that trait (SI Appendix, section S20). These pairings may have lowered quality partly by making the AI more compliant and sycophantic and inspiring it to push back less on its human collaborators.

AI experience moderated how AI personality affected ad quality, while employment status and country of origin showed additional exploratory heterogeneity (SI Appendix, sections S8 and S17). Taken together, these results suggest the existence of economically and statistically significant gains from AI personalization. When people collaborate with AI with personalities that are more appropriate for them, the quality of their output improves significantly.

\subsection*{Personality Pairing and Teamwork}

Personality pairings shaped teamwork more weakly than ad quality (preregistered; see SI Appendix, section S17). Participants completed a 35-item teamwork quality survey after the task \citep{hoegl2001}, which assessed six facets of teamwork grouped into a social dimension (mutual support, effort, and cohesion) and a task dimension (communication, coordination, and balance of member contributions). No individual personality-pairing interactions survived BH correction, and neither the social nor the task dimension showed joint significance. The largest social-teamwork coefficients reached about 0.20 SD (SI Appendix, section S13), consistent with small pairing effects that our sample was underpowered to resolve. See SI Appendix, section S7, for the full coefficient plots and regression tables.

\subsection*{Ad Performance}

\begin{figure}[!t]
\centering
\includegraphics[width=0.55\textwidth]{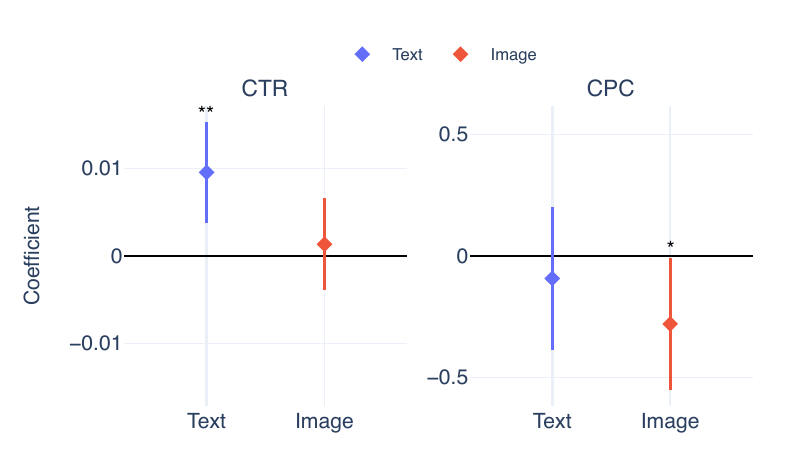}
\caption{Effect of ad quality on field performance. Mixed-effects regression coefficients with campaign random effects for the association between human-rated ad quality (text, image) and click-through rate (CTR) or cost-per-click (CPC); symbols reflect raw $P$-values from the quality regression, which adjusts for realized spend. Symbols indicate significance: $^{\dagger}P < 0.10$, $^{*}P < 0.05$, $^{**}P < 0.01$, and $^{***}P < 0.001$. Error bars show 95\% CIs. See SI Appendix, section S7, for the full field regression tables, and SI Appendix, section S12 for BH corrections across all testing families.}
\label{fig:field_performance}
\end{figure}

Having established the effects of personality pairing on human-rated ad quality, we evaluated the real-world performance of the ads in a large-scale field experiment on X \citep{braun2024adplatforms, braun2024abtesting}. The field experiment measured click-through rates (CTR) and cost-per-click (CPC). The quality of the ads predicted their field performance. A one-standard-deviation increase in human-rated text quality predicted a 7.0\% increase in CTR. A one-standard-deviation increase in human-rated image quality predicted a 3.9\% (or \$0.30) reduction in CPC (Fig.~\ref{fig:field_performance}). These quality--performance associations adjust for realized spend because platform delivery can allocate spend unevenly across ads.

\section*{Discussion}

Our study provides large-scale experimental evidence that personality pairing improves human--AI collaboration and team performance and highlights the need to treat AI agents not just as tools, but as teammates whose traits can be intentionally engineered for collaborative success. Recently, others have tailored human approaches to working with AI, for example by shaping how salespeople interact with AI under different decision-making authority, training, and incentives conditions, to improve sales performance \cite{krakowski2025human}. We measure the performance effects of AI personalization in a context in which AI personalities are directly and randomly manipulated. As AI agents become more prevalent in collaborative settings, designing them with personality traits that complement their human counterparts could enable optimization of performance in real agentic workflows and support more effective and satisfying collaborations.

Our results challenge simple accounts of person-team fit in human--AI collaboration. In human teams, person-team fit theory distinguishes supplementary fit, based on similarity, from complementary fit, in which partners fill each other's gaps \citep{barrick1998gma, bell2007meta, kristofbrown2005consequences}. Applied to human--AI pairs, these accounts would imply that effective pairing should follow from either matching similar personalities or combining different strengths. The evidence points to a more specific form of fit. Some different-trait pairings performed poorly, including extraverted humans with conscientious AI, conscientious humans with agreeable AI, and neurotic humans with conscientious AI. Other pairings were neutral or beneficial. Similarity and difference are therefore too coarse to explain which pairings help or hurt. Effects instead depend on the particular personality pairing and how that combination shapes the collaboration.

The findings point to practical pairing implications. For ad quality, extraverted humans produced lower-quality ads when paired with conscientious AI (the strongest effect in the study), conscientious humans produced lower-quality ads when paired with agreeable AI, and neurotic humans produced lower-quality ads when paired with conscientious AI.

The estimated magnitudes are economically meaningful in this advertising setting. In the field experiment, a one-standard-deviation increase in human-rated text quality predicted a 7.0\% increase in CTR, and a one-standard-deviation increase in human-rated image quality predicted a 3.9\% (or \$0.30) reduction in CPC. A single line of personality prompting shifted the largest composite ad-quality result by approximately 0.39 SDs. Given these effects, the pairing effect on quality translates to a \mytilde2.7\% improvement in CTR ($0.39 \times 7.0\%$) and a \mytilde1.5\% improvement in CPC ($0.39 \times 3.9\%$, or about \$0.12). At advertising scale, changes in CTR and CPC of this magnitude materially affect campaign performance.

While ours is the largest randomized controlled trial of AI personalization to date, several limitations circumscribe our conclusions. First, the specific pairing effects we identified in a marketing context may not extend to other collaborative tasks such as coding or analytics. More work is needed to understand AI personalization across tasks, especially given the jagged frontier of AI capabilities, which could simultaneously drive and interact with personality effects.

Second, the pairings we identify were not fully predictable ex ante. Given the diversity of both humans and AI agents, it is critical to experimentally vary and benchmark how humans work with different personalities to find ones that improve collaboration. Our experimental results establish that personality pairing can matter for human--AI collaboration and field performance, and they motivate tuning AI personalities to human collaborators.

Third, our results were obtained with a specific model snapshot (\texttt{gpt-4o-2024-08-06}). Different architectures and alignment procedures produce different behavioral responses to personality prompts, so the specific coefficients are likely model-dependent. The general finding that personality pairing matters may extend to other models capable of prompt-induced behavioral variation. Steering techniques themselves generalize across both models and methods, ranging from prompting to representation-level interventions \citep{zou2023representation}. Diversifying AI personas through personality prompts has been shown to eliminate the homogenization of creative outputs observed when all users interact with identically prompted AI \citep{wan2025diverse}. These prior results suggest that personality-based personalization is a promising design lever rather than an artifact of any single model or technique.

Fourth, the Big Five personality traits may overlook other traits more relevant to human--AI dynamics. Our exploratory analyses point to further heterogeneity in demographic moderators like country of origin and experience with AI (SI Appendix, section S8), and linguistic patterns in collaboration offer another promising direction for future work on dimensions relevant to AI personalization \citep{wan2025personalized}.

Fifth, our scope limits what we can say about longer time horizons, larger teams, and psychosocial effects. A single 40-min session cannot address how pairings play out over repeated interactions, including whether prolonged use produces overreliance on AI \citep{fang2025psychosocial}. We focus on one-to-one pairings rather than multiperson teams or networks of collaborators, where AI could vary its personality across teammates or over time. We measure self-assessed collaboration dynamics with teamwork variables, but we examine no measures of well-being or adverse psychosocial outcomes. As the field of AI personalization develops, it will be important to understand the psychological and social implications of tailoring AI personalities to their human users, and we encourage future research in this area.

While acknowledging these limitations, we believe this large-scale experimental evaluation of personality pairing in human--AI collaboration is an important step toward robust AI personalization. The effects shown here underscore the critical role of personality pairing in improving human--AI collaboration. Specific pairings improved ad quality while others hindered it, and ad quality predicted real-world ad performance. These findings suggest that designing AI agents to be socially and behaviorally adaptive may help them complement human collaborators more effectively. Ultimately, our work demonstrates that strategically pairing AI personalities with human traits has the potential to transform collaborative workflows and performance in human--AI teams.

\section*{Materials and Methods}

This research was preregistered and approved by the MIT Committee on the Use of Humans as Experimental Subjects (see \href{https://osf.io/jfzha}{osf.io/jfzha} and \href{https://osf.io/95dhu}{osf.io/95dhu}). Informed consent was obtained from all participants prior to their participation in the study.

\subsection*{The Pairit Platform}

\begin{figure}[ht]
    \centering
    \includegraphics[width=0.7\linewidth]{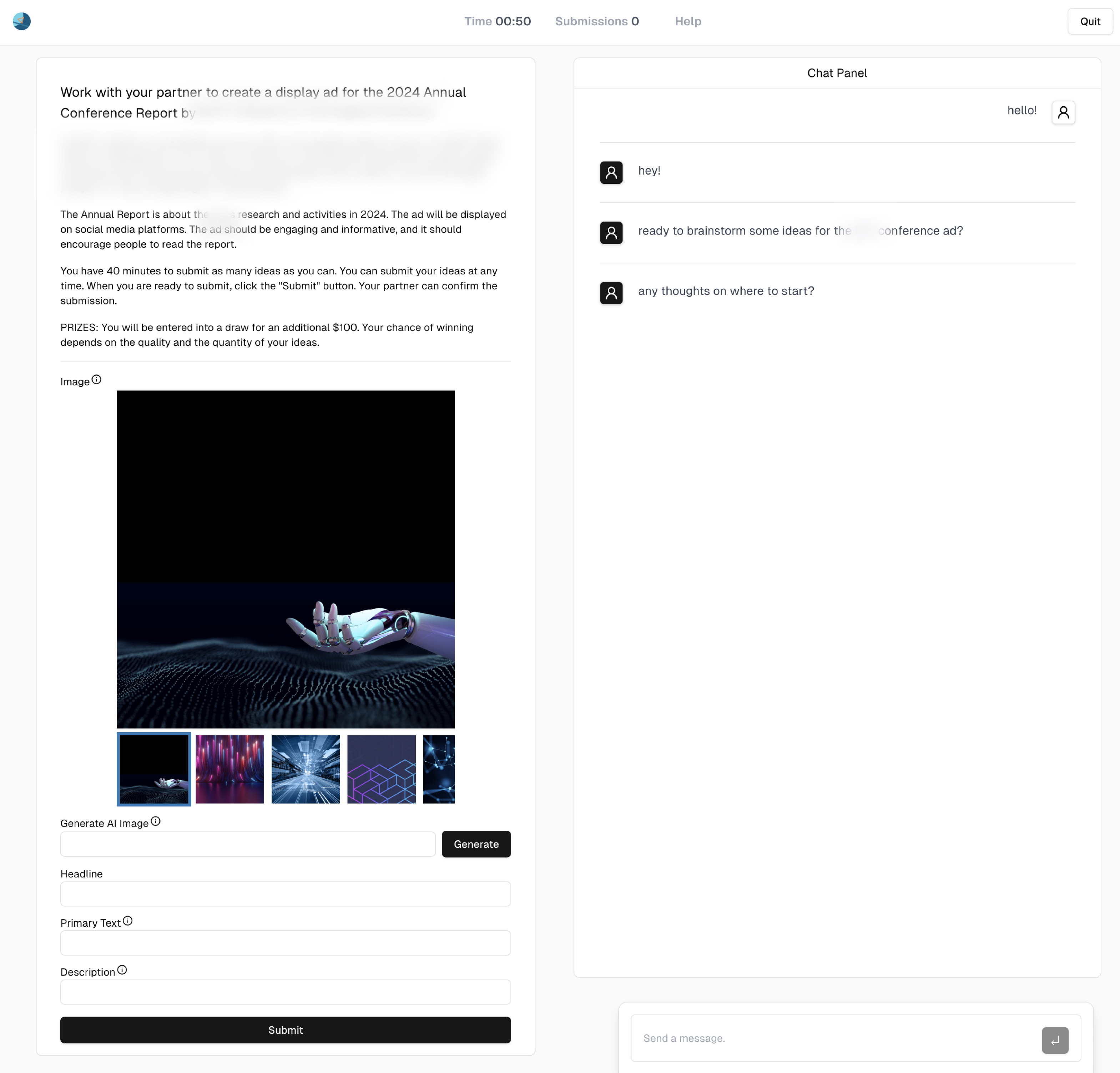}
    \caption{The Pairit platform. The task workspace is on the Left, and the messaging interface on the Right. The participant chats with an AI agent which has full context of the user interface and can modify all elements on the task panel. Some text in the screenshot is intentionally blurred to anonymize identifying information about the partner think tank, and a larger version of the same screenshot is available in SI Appendix, section S4.}
    \label{fig:pairit}
\end{figure}

The study was conducted on the Pairit platform (Fig.~\ref{fig:pairit}). On the Left panel of the platform, users can read the task, select images from a carousel, generate images using Dall-E 3, and write copy for the display ads. In the human--AI teams, the AI has full context of all elements on the platform, including screenshots of the ad images as they evolve over time during the collaboration. In all conditions, only the participants can submit ads. Additional details on the platform can be found in SI Appendix.

To enable the AI agents to act as effective collaborators on the Pairit platform, we provided them with full context of the task and collaboration by prompting each LLM call with all information on the user interface. The prompts included the same text of the task given to participants, previous submissions, personality prompts, current copy, elapsed time, history of its own actions, history of its chain-of-thought, chat history, general instructions, and a screenshot of the image after each change by the participant. We used OpenAI's multimodal \texttt{gpt-4o} model (specifically the \texttt{gpt-4o-2024-08-06} snapshot) with a temperature of 0.2 and default nucleus sampling parameters; no top-$p$, max token, or seed constraints were imposed, so sampling was stochastic. Image generation used \texttt{dall-e-3} at $1{,}024 \times 1{,}024$ resolution. Additional details on the AI agent are in SI Appendix, section S2.

Using Pairit, we designed the AI agent to be able to take the same actions that a human participant could take, apart from submitting ads. The actions include sending chat messages, editing any element of the ad copy (including the headline, primary text, and description), selecting images, generating images using the Dall-E 3 interface, or waiting (i.e., not taking any action). The agent is prompted every 10 s on whether to engage to take an action.

To ensure the agent is taking actions appropriately, we use chain-of-thought prompting using OpenAI's structured output feature \citep{wei2023chainofthought}. Specifically, we prompt the model with questions to reflect on the state of the collaboration. These questions were necessary to avoid undesirable model behavior, such as repeatedly sending the same message.

\subsection*{Personality Prompts}

We note that these prompts induce behavioral styles rather than latent psychological traits; we use ``personality'' throughout as shorthand for prompt-induced behavioral tendencies.

\begin{figure}[ht]
    \centering
    \includegraphics[width=0.7\linewidth]{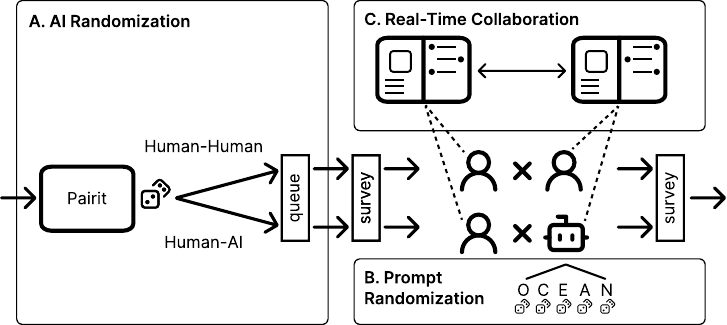}
    \caption{
        The experimental design for investigating human--AI collaboration. (A) Participants are randomized into collaborating with either a human partner or an AI agent. (B) AI agents are randomly assigned prompts to induce low or high levels of the Big Five personality traits. (C) Pairs collaborate in a real-time collaborative workspace to create display ads.
    }
    \label{fig:overview}
\end{figure}

For the randomization of AI personalities, we prompt the model with prompts that induce high or low levels of the Big Five personality traits. Each of the five traits is independently randomized to be high ($P = 0.5$) or low ($P = 0.5$; Fig.~\ref{fig:overview}B) using P\textsuperscript{2} prompting, which generates a detailed description of individuals with the traits \citep{jiang2023evaluating}. This factorial design means that the average effect of a prompted AI trait, such as high versus low neuroticism, averages over all possible assignments of the other four AI traits. The interaction terms then estimate whether that high-versus-low AI-trait contrast differs by the human partner's measured trait score. Since human personality is measured rather than randomized, these interaction estimates are causal with respect to the randomized AI-trait assignment and descriptive with respect to variation in human traits. See SI Appendix, section S5, for sample summary statistics, balance across the human--AI and human--human conditions, and the independent AI-trait randomization design.

We validated that the personality prompts produced measurable behavioral differences in AI messages. For each trait, we extracted a prespecified set of linguistic features from AI-generated messages and constructed composite behavioral indices using the sum of standardized feature values with theoretically expected signs \citep{anderson2008multiple}. Examples of core behavioral markers include: 1) Openness: emoji usage, conservative language (reversed), and exploratory suggestions; 2) Conscientiousness: first-message elaboration (length and word count); 3) Extraversion: enthusiasm markers, exclamation usage, and words per message; 4) Agreeableness: permission-asking ratio, affirming language, and we/our orientation; and 5) Neuroticism: cautious language, fewer alternatives offered (reversed), and less confident language (reversed).

All five composites distinguished their target trait condition (high versus low) at $P < 0.05$, with Cohen's $d$ ranging from 0.12 to 0.17 (SI Appendix, section S6). Agreeableness had the strongest individual behavioral feature (permission-asking; $d = 0.14$, $P = 0.013$), and conscientiousness had the broadest manipulation (two individually significant features). A discriminant validity matrix confirmed that each composite preferentially differentiated its own trait condition, with the largest effect on the diagonal in four of five cases. LLMs can exhibit social desirability biases that attenuate personality manipulations \citep{salecha2024social}, which may partly explain the modest effect sizes of our behavioral composites.

\subsection*{Procedure} \label{sec:methods:procedure}
As soon as participants were redirected to our platform from Prolific, they were randomized into either the human--human or the human--AI condition (Fig.~\ref{fig:overview}A). In the human--human condition, which we do not discuss in this study, participants joined a queue until another participant was available, at which point they were paired with each other. In the human--AI condition, which is the focus of this study, a participant joined a simulated queue, in which they waited for a random amount of time between 1 and 5 s, after which they were paired with an AI agent. We did not reveal whether or not the partner is a human or an AI until the posttask survey. Before the participants entered the task, they answered a 10-item survey, on a 7-point Likert scale, to measure their Big Five personality traits \citep{rammstedt2007personality}. See SI Appendix, section S5, for summary statistics and covariate balance of the human--AI randomization, and SI Appendix, section S3, for the rating interface and ad mockups used in the evaluation study.

Following the pairing and completion of the pretask survey, participants entered a collaborative workspace where they could chat and collaboratively develop and submit display ads (Fig.~\ref{fig:overview}C). Participants were allotted 40 min to create and submit as many display ads as possible. After 40 min, participants were automatically redirected to a posttask survey.

Upon completion of the task, participants completed a 35-item survey assessing teamwork quality and perception of AI \citep{hoegl2001}. The original instrument included 38 questions on a 7-point Likert scale, but three communication-related items were excluded as they were irrelevant to a one-time online collaborative effort. Next, the survey included two items on prior AI experience (i.e., ``I have previously used AI chatbots (e.g., ChatGPT, Bard),'' and ``My experiences with AI chatbots have been positive''), one item on whether they perceived their partner as AI (i.e., ``I thought my collaborator was an AI during the task''), and a final item disclosing the partner's identity (human or AI) and asking if this altered their view of the collaboration quality (i.e., ``Your collaborator was an AI/a human. How has this knowledge affected your perception of the collaboration's quality?''). 85\% of participants completed the posttask survey.

\subsection*{Participants}

We preregistered a target sample of 2,500 individuals recruited via Prolific (\href{https://prolific.com/}{prolific.com}) located in the United States, with balanced representation by gender and ethnicity. Ultimately, 2,310 participants completed the task. Of the 2,500 individuals who entered, 23 failed to join the queue and were excluded from the study. Additionally, 167 participants either withdrew or exceeded the time limit prior to being paired with a collaborator. The overall attrition rate was 7.6\%. The study was conducted from October 15 to 18, 2024, and took a median of 46.32 min to complete. The participants were paid \$9.

\subsection*{Incentives}

To incentivize the production of high-quality ads, we informed the participants of their eligibility for an additional \$100 bonus payment, determined by the volume and quality of their submissions. Participants were explicitly instructed that ``the greater the number of ads, the greater your chances, but not if the ads are of low quality.'' Ultimately, two participants were awarded \$100 each.

\subsection*{Human Evaluation of Ad Quality} \label{sec:methods:human_eval}

To obtain human ratings of ad quality, we recruited a separate sample of 1,168 raters from Prolific based in the United States, with representative stratification across gender, age, and ethnicity. We built a custom platform for this survey, run on Google Cloud Platform's App Engine. The code is available on GitHub. This evaluation was conducted from November 7 to 9, 2024, and took a median of 16.47 min to complete.

To obtain ratings for all ads while avoiding survey fatigue, we created a random sample, with random order, of 40 ads per participant. To ensure that each ad received at least 3 ratings, we produced a set of 1,300 samples. As a participant entered our survey platform, we drew one sample from the set, without replacement, to provide to the participant, with each participant receiving a unique sample.

To estimate ad performance, we created and used mockups of display ads in the survey (SI Appendix, section S3). For each ad, we asked participants to provide three ratings regarding the quality of the text, the image, and the estimated click through rate. Each item was rated on a 7-point Likert scale. The first rating was based on the following prompt ``The text is present, clear, relevant, and engaging;'' the second on ``The image is visually appealing;'' and the third on ``I am likely to click on this ad.'' We construct the ad quality index at the individual rating level as the unweighted mean of each rater's three 7-point evaluations of a given ad: text quality, image quality, and click likelihood.

\subsection*{Field Evaluation of Ad Performance}

We evaluated real-world ad performance on the X platform, preregistered at \href{https://osf.io/95dhu}{osf.io/95dhu}. Key metrics included click-through rate (CTR), cost-per-click (CPC), view-through rate (VTR; fraction of document viewed), and view-through duration (VTD; seconds). VTR and VTD were tracked via a unique DocSend link for each ad. Ad platforms like X use autobidding algorithms that optimize delivery based on expected engagement, producing divergent targeting across ads within a campaign \citep{braun2024abtesting}. Following Braun et al.\ \citep{braun2024adplatforms}, we estimate causal effects of ads conditional on the platform's optimization algorithm. The relevant estimand is the effect of ad content on performance given the audience the platform selected for each ad, which is the economically relevant quantity for advertisers operating within algorithmic environments. Viewer composition is therefore not constant across ads within a campaign, but campaign random effects absorb systematic differences in audience pools across campaigns. ZIP-code-level randomization across campaigns prevents audience spillovers. No holdout group was used, as the unique DocSend links imply zero visits without ads \citep{braun2024adplatforms}.

The field experiment used a subset of the human--AI ads. We sampled 1,152 ads from the 7,266 ads produced in the human--AI lab condition using stratified random sampling based on the predicted click likelihood implied by the human ratings. We divided the full human--AI ad pool into 10 click-score strata spanning the ad quality distribution and sampled within those strata, rather than stratifying separately within each AI personality condition, so the selected field sample can reflect both the randomized AI traits and the quality-based sampling rule. Balance checks in SI Appendix show that the selected field sample remains balanced across the five AI traits, and AI traits do not jointly predict field-sample inclusion (SI Appendix, section S15). These ads were then deployed, alongside human--human ads, in 400 geographically nonoverlapping campaigns, each with a five-ad split test, for 2,000 ads in total. The selected ads covered 51.7\% of human--AI lab participants (651 of 1,258), and each included participant contributed a mean of 1.77 ads. Eight sampled ads were excluded for potential policy violations on the X platform (e.g., violence, sexual, or drug content). Each campaign targeted 133 random ZIP codes with populations between 10,000 and 100,000, as per the \href{https://data.census.gov/table/ACSDT5Y2020.B01003}{2020 American Community Survey}. One-way ANOVA yielded nonsignificant results across ZIP codes: $F(399, 53199) = 0.954$, $P = 0.734$ for population; $F(399, 53199) = 0.973$, $P = 0.636$ for income. Ads were run from January 21 to February 9, 2025, with 50 campaigns active for two days each due to platform constraints. Campaign random effects controlled for temporal confounders.

\subsection*{Multiple Testing Corrections}

Our experimental design yields 25 human--AI personality interaction terms per outcome (5 human traits $\times$ 5 AI traits). To account for multiple testing, we group related interaction tests into coherent outcome families and apply BH corrections within each family. These families cover ad quality, teamwork, productivity (reported in SI Appendix), and field performance, with a separate correction for each field outcome (CTR, CPC, VTR, VTD). We use the BH correction framework because our objective is to identify a set of statistically supported findings while controlling the expected false discovery rate among reported results, rather than to minimize the probability of a single false rejection across the entire family of tests. Given the number of related hypotheses evaluated, a correction framework akin to Bonferroni would impose substantially lower power and materially increase the risk of false negatives. Since our study is intended to characterize the pattern of supported effects rather than to certify that the probability of any false rejection is below a fixed threshold, BH is the most appropriate framework. We report both raw and corrected $P$-values. The preregistration mapping table in SI Appendix, section S17, indicates which analyses are treated as preregistered versus exploratory.

\subsection*{Model Specifications}
We use outcome-appropriate regression models for human--AI team outcomes. For ad quality and teamwork (7-point Likert scales), we estimate ordinal logit models to respect the ordinal nature of the outcome. For productivity (submission count), we estimate Poisson generalized linear models with heteroskedasticity-robust (HC1) SEs to accommodate dispersion in the count distribution. OLS results are reported in SI Appendix as a robustness check and are substantively similar. The general specification is
\begin{equation}
    g(E[Y_{i}]) = \sum_{h\in \text{BF}}\beta_{h} T_{h,i} + \sum_{a\in \text{BF}} \beta_{a} T_{a,i} + \sum_{h\in \text{BF}}\sum_{a\in \text{BF}} \beta_{ha} T_{h,i} T_{a,i}
    \text{,}
    \numberthis \label{eq:personality}
\end{equation}
where $g(\cdot)$ is the link function (identity for OLS, log for Poisson, logit for ordinal), $Y_{i}$ is the outcome for observation $i$, $\text{BF}$ are the Big Five traits, $h\in \text{BF}$ is a human trait, $a \in \text{BF}$ is a prompted AI trait, $\beta_h$ and $\beta_a$ are coefficients for the main effects of human and AI traits respectively, $\beta_{ha}$ captures the interaction between human trait $h$ and AI trait $a$, $T_{h,i}$ is the normalized (0 to 1) score on trait $h$ for the human participant associated with observation $i$, $T_{a,i}$ is an indicator equal to 1 if the AI partner was prompted to exhibit high levels of trait $a$ (0 for low). The unit of observation $i$ varies by outcome: For ad quality ratings, it is the individual rating; for productivity, teamwork quality measures, and message type fractions, it is the individual participant (with heteroskedasticity-robust SEs). All regressions are estimated within the human--AI condition, so the human--human condition is not part of the estimation sample. Each interaction coefficient $\beta_{ha}$ in Fig.~\ref{fig:personality} therefore captures how the association between human trait $h$ and the outcome differs when the AI partner is assigned high rather than low levels of trait $a$, conditional on the human and AI main effects and the other regressors. Because each AI trait was independently randomized, the effect of one AI trait averages over the other $2^{4}$ possible AI-trait combinations. For example, if $\beta_{ha} = 0.5$ for Conscientiousness$_{\text{H}}$ $\times$ Conscientiousness$_{\text{AI}}$, then moving from the minimum to the maximum normalized human conscientiousness score increases the model's linear predictor by $0.5$ more when paired with high-conscientiousness AI than with low-conscientiousness AI.

To relate human-rated quality to field performance (Fig.~\ref{fig:field_performance}), we regress each field outcome on the human-rated text and image quality with a campaign random effect, adjusting for realized advertising spend because platform delivery can allocate spend unevenly across ads:
\begin{align} \label{eq:field}
    Y_{ic} = \beta_{\text{text}} T^{\text{text}}_{i} + \beta_{\text{image}} T^{\text{image}}_{i} + \gamma \text{Spend}_{ic} + u_c + \epsilon_{ic}
    \text{,}
\end{align}
where $Y_{ic}$ is the field outcome (CTR or CPC) for ad $i$ in campaign $c$, $T^{\text{text}}_{i}$ and $T^{\text{image}}_{i}$ are the human-rated text and image quality, $\text{Spend}_{ic}$ is realized advertising spend, and $u_c \sim N(0, \sigma^2_u)$ is the random intercept for campaign $c$. We also estimate the preregistered personality-pairing field model, using Eq.~\ref{eq:personality} with a campaign random effect; full results are reported in SI Appendix, section S7. Marginal spend differences across AI-trait conditions were small and statistically insignificant (SI Appendix, Table S36).

\subsection*{Data, Materials, and Software Availability}
Anonymized study data have been deposited in OSF (\href{https://osf.io/qr5ae}{osf.io/qr5ae}) \citep{ju2026osf}. Analysis code is available at \href{https://github.com/harangju/personality-pairing}{github.com/harangju/personality-pairing} \citep{ju2026code}.

\section*{Acknowledgments}
We thank Dean Eckles, John Horton, and members of the Massachusetts Institute of Technology (MIT) Social Analytics Lab for their invaluable discussions. This work was supported by the MIT Initiative on the Digital Economy at the MIT Sloan School of Management and was approved by the Massachusetts Institute of Technology Institutional Review Board.

\noindent\textit{Author contributions:} H.J. and S.A. designed research; performed research; analyzed data; and wrote the paper.

\noindent\textit{Competing interest statement:} The authors cofounded a company, called Pairium, to fund and distribute the Pairit experimentation platform for profit and for free to academic researchers for research purposes. The company also provides AI personalization services.

\bibliography{refs}

\end{document}